\documentclass[prc,amsfonts,reprint, superscriptaddress,showpacs,
notitlepage,nofootinbib,longbibliography
]{revtex4-1} 
\pdfoutput=1
\usepackage{hyperref}
\hypersetup{colorlinks=true, linkcolor=blue, citecolor=red, urlcolor=blue}
\usepackage{tikz}
\usetikzlibrary{arrows}
\usetikzlibrary{decorations.pathmorphing}
\usepackage{hyperref}
\usepackage{amssymb}
\usepackage{amsmath}
\usepackage{color}
\usepackage{xcolor}
\usepackage{graphicx}
\usepackage{soul}
\usepackage{mathtools}
\usepackage{braket}
\usepackage{float}

\newcommand{\be}{\begin{equation}\begin{aligned}\label}{}
\newcommand{\ee}{\end{aligned}\end{equation}}
{}
{}
{}

\newcommand{\ii}{{\rm i}}

\newcommand{\bes}{\begin{equation}\small\begin{aligned}}{}
\newcommand{\comment}[1]{}
\DeclareMathSizes{10}{9}{6}{6}

\begin{document}

\title{Thermal rectification in a qubit-resonator system }

\author{L. Magazz\`u}
\affiliation{Pico group, QTF Centre of Excellence, Department of Applied Physics, Aalto University School of Science, P.O. Box 13500, 00076, Aalto, Finland}
\affiliation{Institute for Theoretical Physics, University of Regensburg, 93040 Regensburg, Germany}

\author{E. Paladino}
\affiliation{Dipartimento di Fisica e Astronomia Ettore Majorana, Universit\`a di Catania, Via S. Sofia 64, I-95123, Catania, Italy}
\affiliation{INFN, Sez. Catania, I-95123, Catania, Italy; and CNR-IMM, Via S. Sofia 64, I-95123, Catania, Italy}

\author{J. P. Pekola}
\affiliation{Pico group, QTF Centre of Excellence, Department of Applied Physics, Aalto University School of Science, P.O. Box 13500, 00076, Aalto, Finland}

\author{M. Grifoni}
\affiliation{Institute for Theoretical Physics, University of Regensburg, 93040 Regensburg, Germany}

\date{\today}

\begin{abstract}

A qubit-oscillator junction connecting as a series two bosonic heat baths at different temperatures can display heat valve and diode effects. 
In particular, the rectification can change in magnitude and even in sign, implying an inversion of the preferential direction for the heat current with respect to the temperature bias. We perform a systematic study of these effects in a circuit QED model of qubit-oscillator system and find that the features of current and rectification crucially depend on the qubit-oscillator coupling. While at small coupling, transport occurs via a resonant mechanism between the sub-systems, in the ultrastrong coupling regime the junction is a unique, highly hybridized system and the current becomes largely insensitive to the detuning. Correspondingly, the rectification undergoes a change of sign. In the nonlinear transport regime, the coupling strength determines whether the current scales sub- or super-linearly with the temperature bias and whether the rectification, which increases in magnitude with the bias, is positive or negative.
We also find that steady-state coherence largely suppresses the current and enhances rectification.
An insight on these behaviors with respect to changes in the system parameters is provided by analytical approximate formulas.

\end{abstract}

\maketitle

\section{Introduction}

A heat transport setup, where two heat baths at different temperatures are connected by a junction, displays rectification if the heat current changes in magnitude upon reversal of the temperature bias. Namely the setup has a preferential direction for the heat flow: A device with this property operates as a so-called heat diode~\cite{Pereira2019}. In the context of quantum transport, the simplest model displaying such effect is a two-level system (TLS) connected to bosonic heat baths, namely the spin-boson model~\cite{Leggett1987}.   
In such system, it was found that the current is larger in the configuration where the junction couples more strongly to the cold bath~\cite{Segal2005PRL}. Sufficient conditions for attaining thermal rectification in bosonic quantum transport with a generic junction were laid out in~\cite{Segal2006}: These are the presence of an asymmetric coupling to the heat reservoirs together with the nonlinearity of the junction. However, even for a system with symmetric coupling to the baths, a difference  in the statistics or coherence properties between the baths can produce rectification~\cite{Palafox2022}.
For example, in~\cite{Poulsen2022PRE} it was shown that, 
if one of the baths is equipped with a dark state, under appropriate conditions the current is suppressed
only in the temperature bias configuration where this bath is the one at lower temperature.\\
\indent The role of nonlinearity in heat rectification was explored in detail in~\cite{Bhandari2021} in a model of nonlinear resonator. The large nonlinearity limit recovers the case of a two-level system junction (qubit) for which upper bounds on rectification at weak system-bath coupling were found. The latter are violated at strong coupling. 
On the other hand, expanding in the anharmonicity parameter allows to evaluate thermal rectification in a chain of anharmonic oscillators with no limitations on the junction/bath couplings or temperature~\cite{Motz2018}. Heat  rectification is impacted by the difference between the damping constants of the heat baths and by the sign of the anharmonicity parameter.\\
In the presence of a multi-component junction, the set of internal coupling strengths determines the transport properties in a nontrivial way.  In a model of circuit QED (cQED) heat transport setup, where the junction is formed of two interacting flux-type qubit inductively coupled to dissipative LC oscillators (the heat baths), the rectification largely increases above a certain value of the qubit-qubit coupling, when significant  entanglement between the qubits sets in~\cite{Iorio2021}. 
For two weakly coupled TLSs, the rectification of the heat current is maximized by matching the TLSs' frequencies and increases by increasing the temperature gap between the heat baths~\cite{Liu2024}.\\
\indent A prominent route for experiments on heat transport in the quantum regime is provided by cQED platforms~\cite{Blais2021}, where coupled superconducting qubits and resonators are connected to normal-metal heat reservoirs upon which a temperature bias is imposed~\cite{Pekola2021}. These platforms allow for a fine control over the junction/baths parameters and ultimately on the heat flow through the junction. For example, a heat valve device based on a superconducting qubit was envisioned in~\cite{Ojanen2008} and realized~\cite{Ronzani2018} which modulates the heat current by tuning the qubit frequency via an applied magnetic flux.\\
\indent The nonlinearity of superconducting qubits and the possibility to arrange the components of the junctions so as to produce an asymmetric coupling to the baths, furthermore render cQED setups natural candidates for operating as thermal diode~\cite{Ruokola2009,Karimi2017,Thomas2019,Diaz2021}. In~\cite{Senior2020} a transmon-type qubit, embedded between superconducting resonator, of different frequencies, was used to realize a magnetic flux-tunable thermal diode. In~\cite{Upadhyay2024}, the large nonlinearity of a flux qubit was exploited to provide a proof-of-concept of microwave diode, experimentally demonstrating heat rectification at the qubit symmetry point.\\
\indent A qubit-oscillator system, described by the quantum Rabi model (QRM), placed as a series between the heat baths constitute an inherently asymmetric and nonlinear junction. As such this system can display heat rectification.
In~\cite{Chen2022}, heat transport and photon statistics were addressed in the QRM with longitudinal and transverse qubit-oscillator coupling finding that the longitudinal component of the coupling is responsible for enhancing the heat current at strong qubit-oscillator coupling. The two-photon correlation function in the resonator exhibits an antibunching-to-bunching transition by tuning the coupling mixing angle. A connection between thermal rectification and the optical emission properties of the resonator in  was put forward~\cite{Liu2023} in an extension of the QRM, where the qubit-oscillator Hamiltonian is quadratic in the oscillator creation/annihilation operators, the so-called two-photon Rabi model. \\
\indent In this work, we explore heat transport and rectification in a model of cQED setup where the junction is formed of a flux-type qubit biased by an applied magnetic flux, coupled to a superconducting resonator. Such realization of the QRM is of experimental relevance as it allows for achieving large coupling strengths, well into the ultrastrong coupling (USC) regime and even beyond~\cite{Yoshihara2017,Forn-Diaz2018review,Kockum2019,Giannelli2024}. 
The system considered is the same as in~\cite{Magazzu2024PRB,Magazzu2025}, where the thermal conductance was addressed down to low temperatures using a beyond-leading-order diagrammatic approach. 
Here, we study the heat current and rectification in the same device,  assuming the junction to be weakly coupled to the heat baths, and the temperature sufficiently high so that a leading order treatment in the junction-bath coupling is justified.  Steady-state coherent effects are accounted for in our treatment. 	\\
\indent Depending on the junction's parameters, namely on the qubit-resonator coupling and detuning, different transport/rectification regimes emerge. The flux qubit can be biased by an applied magnetic flux: At the symmetry point, we find that the rectification undergoes a change of sign by increasing the coupling into the USC regime, namely the preferential direction for the heat flow changes. In this regime, a significant enhancement of rectification occurs when the qubit and the oscillator are \emph{off-resonance}. The presence of a bias on the qubit yields a richer picture: The rectification turns, as a function of the bias, from positive to negative at weak to intermediate coupling and \emph{vice-versa} in the USC regime. As a general feature, the rectification is consistently negative around resonance, independent of the qubit-oscillator coupling. Quasi-degenerate doublets of excited states emerge in the spectrum of the QRM at resonance and weak qubit-oscillator coupling. In this case, nonvanishing steady-state coherences  suppress the heat current and enhance rectification to an extent that depends nontrivially on the system-bath coupling.  
\\
\indent We also study the transport properties as a function of the temperature bias $\Delta T$ for different values of qubit-oscillator detuning and coupling. At resonance, the current grows sub-linearly and the rectification is negative and enhanced at large coupling, whereas off-resonance the current goes from a super-linear to a sub-linear dependence on $\Delta T$  by increasing the coupling and, correspondingly, the rectification turns from positive to negative. Our study highlights the nontrivial and complex behavior of heat transport in the QRM, including coherent effects on current and rectification. 
Finally, we detail a practical implementation of a heat transport setup based on superconducting circuits which can achieve the transport and rectification regimes discussed throughout this work.

\begin{figure}[ht!]
\begin{center}
\includegraphics[width=0.45\textwidth,angle=0]{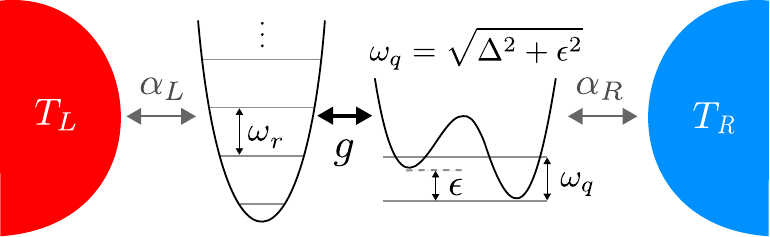}
\caption{\small{Heat transport setup. The system is formed by a flux qubit interacting with a resonator (harmonic oscillator) of frequency $\omega_r$. The qubit frequency is $\omega_q=\sqrt{\Delta^2+\epsilon^2}$, where $\epsilon$ is the qubit bias and $\Delta$ is the qubit frequency splitting at zero bias. The qubit-oscillator interaction strength is $g$. These two elements of the quantum Rabi model are weakly coupled to independent ohmic baths with dimensionless coupling strengths $\alpha_\nu$.}}
\label{fig_setup}
\end{center}
\end{figure}

\section{Setup}
\label{setup}

\subsection{Superconducting qubit-oscillator junction}
\label{sec_setup}

The model of heat transport setup considered in this work is shown in Fig.~\ref{fig_setup}. It is formed of a superconducting artificial atom interacting with a mode of the electromagnetic field, realized by a flux qubit and a superconducting resonator, respectively. The qubit is characterized by the tunneling splitting $\Delta$ and the applied bias $\epsilon$ giving the frequency gap $\omega_q=\sqrt{\Delta^2+\epsilon^2}$. The resonator is modeled as a quantum oscillator of frequency $\omega_r$. This setup is described by the quantum Rabi Hamiltonian which, in the basis of the qubit persistent current states $\{\ket{\circlearrowright },\ket{\circlearrowleft}\}$, has the form~\cite{Forn-Diaz2010,Yoshihara2017}
\begin{equation}
\frac{\hat H_{\rm Rabi}}{\hbar}=-\frac{1}{2}\left(\epsilon\sigma_z + \Delta\sigma_x \right) +\omega_r \hat a^\dagger \hat a + g\sigma_z \left(\hat a^\dagger + \hat a \right)\;,
\label{H_Rabi}
\end{equation}
where $\sigma_z=\ket{\circlearrowright}\bra{\circlearrowright}-\ket{\circlearrowleft}\bra{\circlearrowleft}$ and $\sigma_x= \ket{\circlearrowright}\bra{\circlearrowleft}+\ket{\circlearrowleft}\bra{\circlearrowright}$ and where $a^\dag$ ($a$) creates (annihilates) an excitation of the field mode.  The angular frequency $g$ quantifies the qubit-oscillator coupling strength.\\ 
\indent While the closed quantum Rabi model is characterized by the two frequency scales $\omega_q/\omega_r$ and $g/\omega_r$, in practical realizations interactions with the environment introduce loss mechanisms. If $g$ is larger than the loss rates $\kappa_i$, one observes  damped Rabi oscillations, a signature of the strong coupling regime. The regime $\kappa_i<g\ll\omega_r,\omega_q$ is well described within the rotating wave approximation (RWA), where the so-called counter-rotating coupling terms in the interaction part of the Hamiltonian are neglected, resulting in the Hamiltonian of the Jaynes-Cummings model. The latter is the archetypal model of atom-photon interaction used in quantum optics~\cite{Shore1993}
. 
The nonperturbative USC regime sets in for $g\gtrsim 0.3~\omega_r$~\cite{Forn-Diaz2018review}. In this regime of light-matter interaction, perturbation theory in $g$ fails and the counter-rotating terms have to be taken into account. \\
\indent Several analytical approaches to the quantum Rabi model exist yielding exact~\cite{Braak2011} and approximate expressions for its eigensystem in different parameter regimes~\cite{Irish2007,  Ashhab2010, Hausinger2008, Hausinger2010PRA,LeBoite2020} and even in the presence of dissipation~\cite{Vaaranta2025}. A perturbative treatment beyond the RWA, thus accounting for the effect of the counter-rotating terms in the interaction Hamiltonian, is provided by the second-order Van Vleck perturbation theory ~\cite{Hausinger2008}.  
A  simple approximation scheme which is nonperturbative in the qubit-oscillator coupling is the generalized rotating wave approximation (GRWA), originally developed in~\cite{Irish2007} for zero bias, and generalized in~\cite{Zhang2013} to $\epsilon\neq 0$, see also~\cite{Magazzu2024PRB,Magazzu2025}. This approximation scheme is perturbative in the dressed qubit gap $\tilde{\Delta}_{ij}= \Delta e^{-\tilde\alpha /2}\tilde\alpha^{(i-j)/2}\sqrt{j!/i!}\;\mathsf{L}^{i-j}_j(\tilde\alpha)$ ($i\geq j$), where $\tilde\alpha:=(2g/\omega_r)^2$ and $\mathsf{L}^{k}_n$ are the generalized Laguerre polynomials. Here, we use this approximation scheme to gain insight on the heat transport features when $\Delta<\omega_r$. Introducing the renormalized qubit frequency $\omega_{q,n}:=\;\sqrt{\tilde\Delta_{nn}^2 + \epsilon^2}$,
within the GRWA, the frequency spectrum of the QRM is approximated by 
\begin{equation}
\begin{aligned}\label{eigensys_GRWA}
\omega_0 &=\;-\frac{\omega_{q,0}}{2}- \frac{g^2}{\omega_r}\;,\quad
\omega_{n}^\mp=\;n\omega_r-\frac{\omega_{q,n}}{2} - \frac{g^2}{\omega_r}+\frac{X_n^\mp}{2}\;,
\end{aligned}
\end{equation}
where 
$$X_n^\mp:=\delta_n \mp \sqrt{\delta_n^2+\Omega_{n}^2}\;,\quad \delta_n  :=(\omega_{q,n}+\omega_{q,n-1})/2  -\omega_r\;.$$ Here, $\Omega_{n} :=\;\tilde\Delta_{nn-1}(c^+_n c^+_{n-1} +  c^-_n c^-_{n-1})$, and \hbox{$c^\pm_n :=\;\sqrt{(\omega_{q,n}\pm\epsilon)/2\omega_{q,n}}$}. The corresponding energy eigenstates are
\begin{equation}
\begin{aligned}
\Ket{0}&=\ket{{\Psi_{-,0}}}\;,\quad
\Ket{n_\mp}={\rm u}_n^{\mp}\ket{{\Psi_{-,n-1}}}+{\rm v}_n^{\mp}\ket{{\Psi_{+,n}}}\;.
\end{aligned}
\end{equation}
The states $
\Ket{\Psi_{\pm,j}}=c_\pm^j\Ket{-_z j_-}\pm c_\mp^j\Ket{+_z j_+}$ are superpositions of the displaced states ($\ket{\circlearrowright },\ket{\circlearrowleft}\equiv\ket{\pm_z}$)
$$\ket{\pm_z j_\pm}=\ket{\pm_z}D(\pm g/\omega_r)|j\rangle$$
with the displacement operator defined as $D(x):=\exp[ x(\hat a - \hat a^\dag)]$. 
The amplitudes ${\rm u}^\pm_n$ and ${\rm v}^\pm_n$ read
\begin{equation}
\begin{aligned}\label{GRWA_u_v}
{\rm u}_n^\pm &:=\;\frac{X_n^\pm}{\sqrt{\left(X_n^\pm\right)^2+\Omega_n^2}}\;,\quad
{\rm v}_n^\pm &:=\;\frac{-\Omega_n}{\sqrt{\left(X_n^\pm\right)^2+\Omega_n^2}}\,.
\end{aligned}
\end{equation}
These coefficients are relevant as they enter the matrix elements of the system operators that couple to the baths. 
For small $g/\omega_r$, the GRWA is not equivalent to the RWA. Indeed the latter is valid for arbitrary detuning provided that $g/\omega_r\ll 1$ while the former is perturbative in the renormalized $\Delta$. Nevertheless, for $g/\omega_r\ll 1$, neglecting the terms $\mathcal{O}(\tilde{\alpha})$ and substituting $\tilde\Delta_{nn-1}\simeq \sqrt{n}2g\Delta/\omega_r\rightarrow \sqrt{n}2g\Delta/\omega_q$ the above formulas reproduce the RWA, see Appendix~\ref{sec_RWA}.

\subsection{Quantum heat transport setup}
\label{sec_QHT}

\indent In the heat transport setup shown in Fig.~\ref{fig_setup}, the junction is connected to heat baths, indexed by $l=L,R$ and possibly at different temperatures.
The Hamiltonian of the setup reads
\be{H_Caldeira_Leggett2baths}
\hat H=&\;\tilde{H}_{\rm Rabi}+\sum_{l,j}\hbar\omega_{lj} \hat b_{lj}^\dag \hat b_{lj} - \sum_{l}\hat{Q}_l\hat{B}_l\:.
\ee
The bath operators $\hat{B}_l$ collect coupling to the individual bath modes 
via the bath displacement operators, namely $\hat{B}_l\equiv \sum_j \hbar\lambda_{lj}(\hat b_{lj}+\hat b_{lj}^{\dag})$, according to the Caldeira-Leggett model~\cite{Caldeira1981, Caldeira1983}. The coupling is mediated by the dimensionless system operators $\hat{Q}_L=a+a^\dag$ and $Q_R=\sigma_z$
(in the localized qubit basis of Eq.~\eqref{H_Rabi}). 
At zero qubit bias the GRWA yields 
$$Q_{L01}=2g{\rm u}_1^-/\omega_r+{\rm v}_1^-\quad {\rm and} \quad Q_{R01}=-{\rm u}_1^-\;,$$ 
see Eq.~\eqref{GRWA_u_v}.
The baths and their interaction with the system are collectively described by the spectral density function $G_l(\omega)=\sum_{j=1}^{N}\lambda_{lj}^2\delta(\omega-\omega_{j})$. In the continuum limit, the ohmic case has the low-frequency behavior $G_l(\omega)\sim \alpha_l\omega$, where $\alpha_l$ measures the coupling to bath $l$. Finally,  $\tilde{H}_{\rm Rabi}=\hat H_{\rm Rabi}+\sum_l \mu_l\hat{Q}_l^2$, where $\hat H_{\rm Rabi}$ is the quantum Rabi Hamiltonian, Eq.~\eqref{H_Rabi}, and  $\mu_l=\hbar\int_{0}^{\infty}d\omega[G_l(\omega)/\omega]$ is the bath-induced renormalization of the bare system Hamiltonian, see Appendix~\ref{appendix_CLM} for details. The qubit-resonator circuit is, as a whole, a nonlinear system and, in our setup, it is coupled in an asymmetric way to the two heat baths: This implies that the rectification for the present setup is in general nonzero~\cite{Segal2005PRL,Segal2005, Segal2006, Bhandari2021}. However, since the system-bath interaction is linear, the setup does not display negative differential thermal conduction, namely a decrease in the current upon increasing the temperature bias~\cite{Segal2006}.\\
\indent The heat current to the reservoir $r$, with \hbox{$r=L,R$}, is defined as the expectation value $I_r(t)=\langle\hat I_r\hat\varrho_{\rm tot}(t)\rangle$. Here, the current operator is given by 
\be{}
\hat{I}_r=d\hat{H}_r/dt=\ii[\hat H,\hat{H}_r]\;,
\ee
where $\hat{H}$ is the full Hamiltonian of the setup, $\hat{H}_r$ is the free Hamiltonian of bath $r$, see Eq.~\eqref{H_Caldeira_Leggett2baths}, and $\hat{\varrho}_{\rm tot}$ is the total density matrix whose trace over the heat baths gives the junction's reduced density matrix $\hat{\varrho}$. Let us introduce the \emph{forward} (\emph{backward}) steady-state currents, defined as 
\be{}
I_{+(-)}:&=I_R^\infty(T_L=T_{h(c)},T_R=T_{c(h)})\;,
\ee
where $T_h=T+\Delta T/2$ and $T_c=T-\Delta T/2$ for a positive temperature bias $\Delta T$. The definitions entail $I_+\geq 0$ and $I_-\leq 0$.  
Under the condition that the two heat baths are coupled to the junction in an \emph{asymmetric} fashion~\cite{Segal2005PRL, Segal2006}, the setup displays non-zero rectification, namely a preferential direction for the heat flow. As a quantifier of rectification we choose 
\be{Rdef}
\mathcal{R}=\frac{I_++I_-}{I_+-I_-+\eta/(I_+-I_-)}\;.
\ee
This is a modified version of the one used in~\cite{Bhandari2021,Tesser2022}, in that we introduce an additional term in the denominator with $0<\eta\ll{\rm max}\{I_+-I_-\}$ so as to suppress the rectification when the current is vanishingly small, still leaving it unaffected around the maximum of the current, where $\mathcal{R}$ is most meaningful, see~\cite{Khandelwal2023} for an insightful discussion on the performance of heat rectifiers. Throughout this work we set $\alpha=\alpha_L=\alpha_R$ and $\eta=10^{-5}\alpha (\hbar\omega_r^2)^2$.

\section{Redfield equation approach}
For convenience we introduce the Bohr frequencies $\omega_{nm}:=\omega_n-\omega_n$, where $\tilde H_{\rm Rabi}\ket{n}=\hbar\omega_n\ket{n}$. 
To leading order in the junction-bath coupling, the steady-state reduced density matrix $\hat\varrho^\infty$ is the solution of the Redfield equation
\be{RME}
0=-\ii\omega_{nm}\varrho_{nm}^\infty+\sum_{n'm'}\mathcal{K}^{(2)}_{n m n'm'}\varrho_{n'm'}^\infty\;,
\ee
see~\cite{Magazzu2024PRB} for a derivation based on a diagrammatic unraveling of the kernel of the  Nakajima-Zwanzig equation. 
The Redfield tensor reads
\bes\label{K2_bosons}
&\mathcal{K}_{n m n'm'}^{(2)}=\frac{1}{\hbar^2}\sum_{l }\Bigg\{Q_{l m'm}^*Q_{l n n'}
[W_{l nm'}+W_{l mn'}^*]\\
&-\sum_{k}\Big[Q_{l k n}^*
Q_{l k n'}W_{l k m'}\delta_{m'm}+
Q_{l k m'}^*Q_{l k m}W_{l k n'}^*\delta_{n'n}\Big]
\Bigg\}\;,
\ee
with $W_{l nm}\equiv W_l(\omega_{nm})$ the one-sided Fourier transform of the bath correlation function of bath $l$, namely
\bes
W_{l}&(\omega)=\int_{0}^{\infty}dt\langle \hat{B}_l(t)\hat{B}_l(0)\rangle 
e^{-\ii\omega t}
\;.
\ee
Explicit expressions are provided in  Appendix~\ref{appendix_BCF} for a ohmic-Drude spectral density function. 
The power spectral density of the bath operator $\hat{B}$, which enters the expression for the transition rates between energy levels, see below, is defined as $S_l(\omega):=\int_{-\infty}^{\infty}dt\langle \hat{B}_l(t)\hat{B}_l(0)\rangle 
e^{\ii\omega t}=2{\rm Re}W_{l}(-\omega)$.   Note that no Markovian or secular approximation is invoked in Eq.~\eqref{RME}.
\begin{figure}[ht!]
\begin{center}
\includegraphics[height=0.275\textwidth,angle=0]{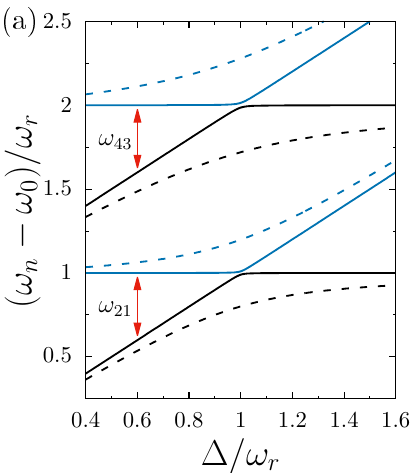}\hspace{-.25cm}
\includegraphics[height=0.275\textwidth,angle=0]{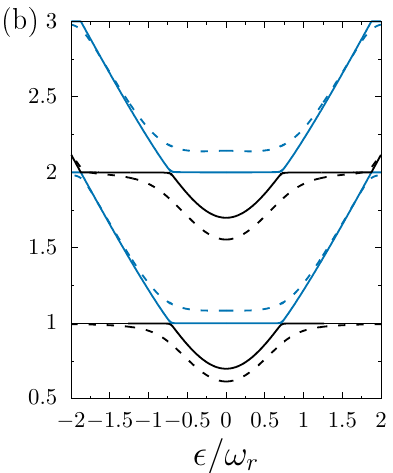}
\caption{\small{Excitation spectrum of the quantum Rabi model vs. the qubit frequency splitting at $\epsilon=0$ (a) and vs. the qubit bias for $\Delta=0.7~\omega_r$ (b). For both panels the qubit-oscillator coupling is $g=0.01~\omega_r$ (solid lines) and $g=0.2~\omega_r$ (dashed lines). The avoided crossings occur at resonance, $\omega_q:=\sqrt{\Delta^2+\epsilon^2}=\omega_r$. }}
\label{fig_spectrum}
\end{center}
\end{figure}
The heat current to bath $R$ reads~\cite{Thingna2012, Magazzu2024PRB}
\be{I2}
I_R^{(2)\infty}
&=-\frac{1}{\hbar^2}2{\rm Re}\sum_{nn'm'}Q_{R m' n}^*Q_{R n n'}\bar{W}_{R n m'}\varrho^\infty_{n'm'}\;,
\ee
where
\be{}
\bar{W}_{lnm}=\;\hbar\omega_{nm}W_{l,nm}+\ii\hbar\langle \hat{B}_l(0)\hat{B}_l(0)\rangle
\;.
\ee
The term with the zero-time correlator $\langle \hat{B}_l(0)\hat{B}_l(0)\rangle$ does not contribute to the current due to the hermiticity of $\varrho^{\infty}$~\cite{Magazzu2024PRB}.\\
\indent  As pointed out in~\cite{Hartmann2020}, within its regime of applicability, the Redfield equation is more accurate than other approaches that enforce positivity and a violation of the latter signals the break-down of weak coupling assumption. In what follows we refer to implementations of Eq.~\eqref{RME} that neglect coherences fully or in part as full secular and partial secular master equation, respectively, not as approximations. The reason is that they  are consistent lowest order treatments of the problem at hand in different parameter regimes. In  Fig.~\ref{fig_spectrum}, the spectrum of the QRM is shown as a function of the qubit-resonator detuning for two values of their coupling $g$.
Let $\gamma$ be a frequency scale characterizing the elements of Redfield tensor. 
At very weak system-bath coupling, $\gamma\ll \omega_{nm}$ for $n\neq m$, the full secular master equation (FSME) is appropriate to describe the Rabi model, even at resonance and weak $g$, namely $\gamma\ll g\ll\omega_r$, a situation in which $\omega_{12},\omega_{34}\ll \omega_{10}$, see Fig.~\ref{fig_spectrum}.\\ 
\indent In the FSME all coherences vanish to lowest order and the steady-state reduced density matrix is given by
\be{BR_full_secular}
\rho_{nm}^{\infty}=&\;0\qquad n\neq m\;,\\
{\rm and}\qquad\;\sum_m\Gamma_{nm} \rho_{mm}^\infty=&0 \;.
\ee
The transition rates $m\rightarrow n$ are defined as
$$\Gamma_{nm}=\sum_l\Gamma^l_{nm}:=\mathcal{K}^{(2)}_{nnmm}\;.$$
Note that probability conservation is satisfied by virtue of the property \hbox{$\Gamma^l_{nn}=-\sum_{m(\neq n)}\Gamma^l_{mn}$}, see Eq.~\eqref{K2_bosons}. Explicitly 
\be{Gamma}
\Gamma^l_{nm(\neq n)}&=\frac{1}{\hbar^2}|Q_{l nm}|^2 S_l(\omega_{mn})
\;.
\ee
In the FSME approach for the heat current, Eq.~\eqref{I2} simplifies to
\be{current_secular}
I_R^{(2)\infty}=\sum_{n,m}\hbar\omega_{mn}\Gamma^R_{nm}\rho_{mm}^\infty\;.
\ee
\indent Consider now the situation where the coupling with the baths is sufficiently large so that $\gamma\sim \omega_{21},\omega_{43}\ll\omega_{10}$. In this case, the partial secular master equation (PSME), which couples the coherences corresponding to these  \emph{small} Bohr frequencies to the populations, is appropriate~\cite{Cattaneo2019,Trushechkin2021,Ivander2022,Magazzu2024PRB}. 
Consistency with perturbation theory to leading order entails
that Eq.~\eqref{RME} splits into
\be{BR_partial_secular}
&0=\omega_{nm}\rho_{nm}^\infty\implies \rho_{nm}^\infty =0\;,\quad {\rm for}\quad |\omega_{nm}|\gg \gamma\;,\\
&0=\;-{\rm i}\omega_{nm}\rho_{nm}^\infty + \sum_{n',m'}\mathcal{K}^{(2)}_{nmn'm'}\rho_{n'm'}^\infty\;\\
&\quad{\rm for}\quad |\omega_{nm}|,|\omega_{n'm'}|\in\{0,\omega_{21},\omega_{43}\}\;.
\ee
In Appendix~\ref{3LSRWA}, the PSME is solved analytically for a three-level truncation of the QRM, with a single small Bohr frequency $\omega_{21}$. Note that, according to the PSME, $\alpha$ is not a mere proportionality factor in the expression for the current, which is in contrast with the  results of the FSME. \\ 
\indent Let us consider now the truncation of the QRM to the two-dimensional subspace $\{\ket{0},\ket{1}\}$. The resulting TLS description of heat transport yields analytical expressions providing some insight on the behavior of the current and rectification in the full QRM.
The forward current $I_+$ to the right bath calculated from Eq.~\eqref{current_secular} yields, for the TLS,
\be{Jp_TLS}
I_R^{(2)\infty}=\frac{\hbar\omega_{10} \gamma^R\gamma^L[n_L(\omega_{10})-n_R(\omega_{10})]}{\gamma^R[1+2n_R(\omega_{10})]+\gamma^L[1+2n_L(\omega_{10})]}\,,
\ee
with $\gamma^l:=2\pi G_l(\omega_{10})  |Q_{l 01}|^2$, see~\cite{Segal2005,Segal2005PRL}. Here, $n_l(\omega)=[\exp(\beta_l \hbar\omega) -1]^{-1}$ is the Bose-Einstein distribution at inverse temperature $\beta_l$.\\
\indent To see how the setup gives rise to rectification, let us consider the proportional coupling configuration
$\gamma^L=\bar\gamma(1-\chi)$ and $\gamma^R=\bar\gamma(1+\chi)$, 
with $\bar\gamma:=(\gamma^L+\gamma^R)/2$.
For identical baths,  $G_L(\omega)=G_R(\omega)$, %
\be{chi}
\chi=\frac{|Q_{R01}|^2 - |Q_{L01}|^2}{|Q_{R01}|^2 + |Q_{L01}|^2}\;.
\ee
We find for the rectification, Eq.~\eqref{Rdef} with $\eta=0$,
\be{R_TLS}
\mathcal{R}=\frac{ \chi[n_L(\omega_{10}) - n_R(\omega_{10})]}{1+n_R(\omega_{10}) + n_L(\omega_{10})}\;,
\ee
see also~\cite{Segal2005PRL}. 
This shows that the rectification vanishes for symmetric coupling
($\chi=0$) and/or vanishing temperature bias. Note that Eq.~\eqref{R_TLS} can give finite rectification for vanishing current, e.g. when one of the matrix elements $Q_{l10}\rightarrow 0$.

\section{Results}
\label{section_results}

\begin{figure}[ht!]
\begin{center}
\includegraphics[width=0.5\textwidth,angle=0]{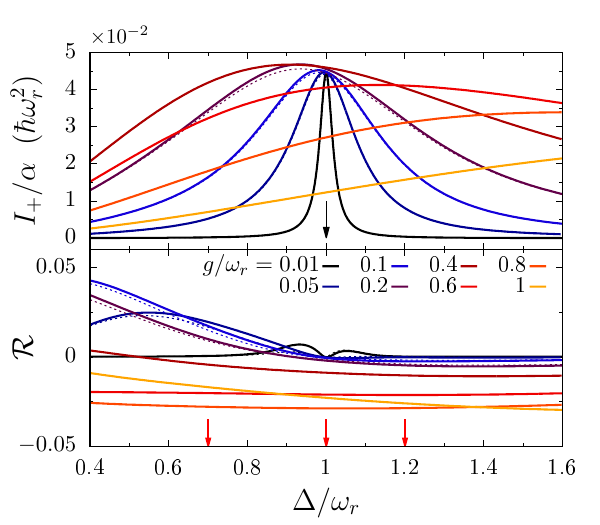}
\caption{\small{Ultrastrong coupling effects in transport at zero bias.
Heat current and rectification \emph{vs.} qubit splitting $\Delta$  at zero bias for different values of the qubit-resonator coupling strength $g$, from weak to ultrastrong. 
The black arrow marks the resonance condition.
Note that $\mathcal{R}$, for $\Delta<\omega_r$, changes from positive to negative upon increasing $g$.
Parameters are $T=0.25~\hbar\omega_r/k_B$,  $\Delta T=0.1~\hbar\omega_r/k_B$, and $\omega_c=5~\omega_r$.
The dashed lines for $g/\omega_r\leq 0.2$ are the analytical results using 2nd-order Van Vleck perturbation theory in $g$, see also~\cite{Magazzu2025}, and the truncation to a four-state system. $\mathcal{R}$ is calculated according to Eq.~\eqref{Rdef} with  $\eta=10^{-5}\alpha (\hbar\omega_r^2)^2$. The red arrows mark, for reference, the selected values of $\Delta$ used in Figs.~\ref{fig_JR_vs_eps} and~\ref{fig_JR_vs_dT} below.}}
\label{fig_JR_vs_Delta}
\end{center}
\end{figure}
In this work, we consider identical baths of the ohmic-Drude type $G_L(\omega)=G_R(\omega)=\alpha\omega/[1+(\omega/\omega_c)]$  with a cutoff at a high frequency $\omega_c$. 
With this choice, analytical expressions for the functions $W_{lnm}$ entering the Redfield tensor, Eq.~\eqref{K2_bosons}, are provided in Appendix~\ref{appendix_BCF}.  
Also, the coefficients $\mu_l$ of the bath-induced renormalization $\sum_l \mu_l\hat{Q}_l^2$ of the system Hamiltonian are explicitly evaluated as $\mu_l=\pi\alpha\hbar\omega_c/2$. Note that this contribution can be safely neglected when the coupling $\alpha$ is very weak, whereas it induces a shift in the heat current at finite coupling, see Appendix~\ref{3LSRWA}. Finally, the temperature bias $\Delta T$ is assumed to be symmetrically applied around the average baths' temperature $T$ as $T_{h/c}=T\pm\Delta T/2$.\\ 
\indent We start in Sec.~\ref{sec_IRvsD} by considering the transport behavior of the setup as a function of the qubit splitting and of the qubit-resonator coupling $g$ at zero qubit bias. In Sec.~\ref{sec_IRvseps}, these two quantities are studied as a function of the applied bias. In Sec.~\ref{sec_IRvsTDT}, we consider how heat transport and rectification behave as a function of the temperature bias for $\epsilon=0$. Finally, in Sec.~\ref{reole_ss_coherences}, the impact of steady state coherences on current and rectification is addressed. Except for Sec.~\ref{reole_ss_coherences}, the results are obtained with the FSME approach, Eqs.~\eqref{BR_full_secular}-\eqref{current_secular}, namely assuming  $\alpha$ to be small enough so that the steady-state coherences vanish to leading order, even in the presence of the avoided crossings at small $g$. This entails that $I_+/\alpha$ is independent of $\alpha$.

\subsection{Heat valve and diode effects at zero qubit bias}
\label{sec_IRvsD}

\indent In Fig.~\ref{fig_JR_vs_Delta}, the forward steady-state heat current $I_+$ and the rectification $\mathcal{R}$ are shown as functions of the qubit frequency splitting $\Delta$ at zero qubit bias, $\epsilon=0$. The resonance condition reads in this case $\Delta/\omega_r=1$.  The curves for $g/\omega_r\leq 0.2$ are reproduced by analytical diagonalization of the QRM within the second order Van Vleck perturbation theory in $g$~\cite{Shavitt1980, Hausinger2008} and truncation to a four-level system, see~\cite{Magazzu2024PRB}.  Perturbation theory in $g$ fails for larger $g$.  
At weak qubit-oscillator coupling, the current is suppressed, except for a resonance peak.
Increasing $g$, the peak broadens and moves towards lower frequencies, $\Delta <\omega_r$. Upon further increasing the coupling, entering the USC regime, the peak further broadens while moving to large frequencies $\Delta>\omega_r$. Correspondingly, for $\Delta/\omega_r<1$, the rectification changes from positive to negative as $g$ is increased, while being always negative at resonance. This change of sign in $\mathcal{R}$ is another transport signature of the transition to the USC.\\
\begin{figure}[ht!]
\begin{center}
\includegraphics[width=0.5\textwidth,angle=0]{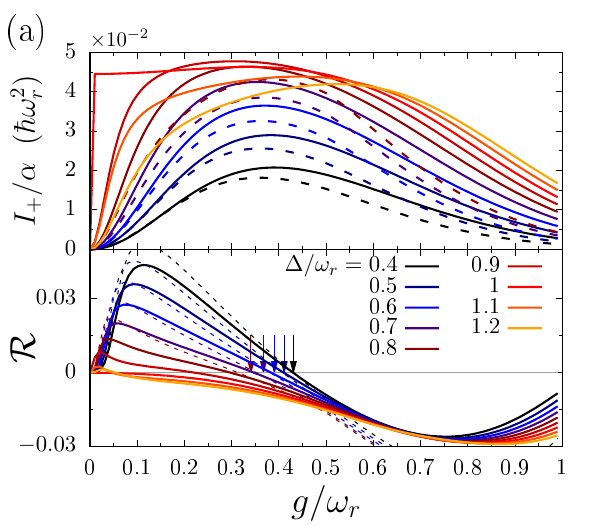}
\includegraphics[width=0.5\textwidth,angle=0]{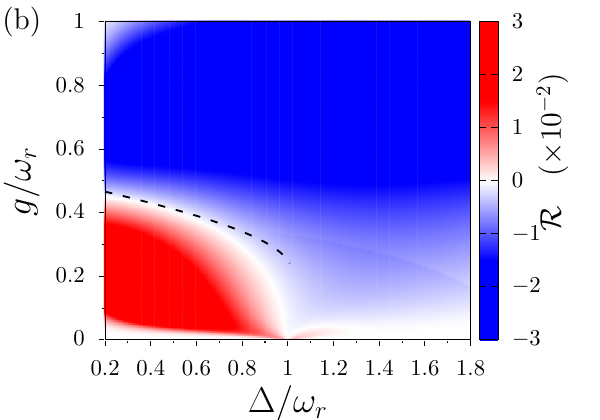}
\caption{\small{Transition to negative rectification at zero bias in the USC.
(a) - Heat current and rectification \emph{vs.} qubit-resonator coupling $g$ at zero bias, $\epsilon=0$ for different values of the qubit splitting $\Delta$. Other parameters are as in Fig.~\ref{fig_JR_vs_Delta}. Note that the rectification at resonance, $\Delta/\omega_r=1$, is always non-positive. The dashed lines for $\Delta/\omega_r\leq 0.8$ are given by the analytical formulas from the TLS truncation, Eqs.~\eqref{Jp_TLS} and ~\eqref{R_TLS}, and GRWA. These approximation capture well the turning point $\mathcal{R}(g^*)=0$. The arrows point at the approximate solution for $g^*$, within the GRWA, given by Eq.~\eqref{gstar}. (b) - Rectification as a function of $\Delta$ and $g$ for the same parameters as in panel (a). The black dashed line marks the condition for zero rectification in a TLS truncation of the model, Eq.~\eqref{chi}, according to the approximate expression~\eqref{gstar}.}}
\label{fig_JR_vs_g}
\end{center}
\end{figure}
\indent The transition from positive to negative rectification upon increasing $g$ is shown in Fig.~\ref{fig_JR_vs_g}(a),  where current and rectification are plotted against the qubit-oscillator coupling at zero qubit bias. Numerical evaluations (solid lines) for the full Rabi model are shown alongside with the corresponding analytical approximate results from applying the GRWA to a TLS truncation of the full Rabi model (dashed lines). The latter scheme captures the turning point of $\mathcal{R}$ for $\Delta<\omega_r$. From this we can establish that rectification vanishes when $|Q_{L01}|=|Q_{R01}|$, see Eq.~\eqref{R_TLS} while, out of resonance, $\Delta\neq\omega_r$, the heat current displays a maximum at an optimal coupling strength $g$ such that $|Q_{L01}|>|Q_{R01}|$, corresponding to negative rectification: This can also be seen by minimizing the current in its TLS expression~\eqref{Jp_TLS}, with respect to $\gamma^R-\gamma^L$.
On the other hand, resonant transport at $\Delta=\omega_r$ occurs for $g/\omega_r\ll 1$, giving large current already at weak qubit-oscillator coupling.\\
\indent  To explore more systematically the behavior of $\mathcal{R}$, in Fig.~\ref{fig_JR_vs_g}(b) the rectification is plotted against both qubit-resonator coupling strength and detuning. The region of weak to intermediate coupling and $\Delta/\omega_r<1$ displays positive rectification while at $\Delta/\omega_r>1$ and/or large coupling the rectification is negative. An intuitive explanation for this behavior can be given by considering two scenarios: In the first, one fixes the coupling in the weak/intermediate regime and moves from negative to positive $\Delta-\omega_r$. In the second, the coupling is increased while keeping $\Delta/\omega_r<1$. In the first case, where the two sub-systems can still be considered as separate objects exchanging excitations sequentially,  the forward direction (where the hot bath is the left one, attached to the resonator) is the preferential one. This is because the level spacing of the resonator is larger and the hot bath populates its excited states that can further decay by exciting the qubit, and so on. Swapping the temperatures, the qubit is populated by the hot bath but it is less likely to populate an excited state of the resonator in order for the latter to decay in the cold bath. A similar reasoning applies, \emph{mutatis mutandis} for $\Delta/\omega_r>1$, where one concludes that the preferential direction for the heat flow is the backward. In the second scenario, one passes from a regime of sequential exchange of excitations between sub-systems to a highly hybridized state where the \emph{dressed} matrix elements of the coupling operators pass from $|Q_{L01}|<|Q_{R01}|$ to $|Q_{L01}|>|Q_{R01}|$ entailing a change in the sign of rectification from positive to negative, see Eq.~\eqref{chi}.\\
\indent As noted above, for $\Delta/\omega_r<1$ the turning point $g=g^*$ where $\mathcal{R}$ changes sign is captured by a TLS truncation of the full QRM, see Fig.~\ref{fig_JR_vs_g}(a). In this approximation, finding the turning point amounts to finding the solution to $\chi=0$, see Eq.~\eqref{chi}, namely solving the equation $|Q_{L01}|-|Q_{R01}|=0$. Within the GRWA, this is equivalent to $|(2g/\omega_r) {\rm u}_1^- + {\rm v}_1^-|= |u_1^-|$, with the coefficients given by Eq.~\eqref{GRWA_u_v}. Using the results of Sec.~\ref{sec_setup} and noticing that the crossing points in Fig.~\ref{fig_JR_vs_g} occur at $g/\omega_r\lesssim 0.5$, we can simplify the equation for $g^*$ at small $\Delta/\omega_r$ to

\be{}
\frac{\sqrt{\tilde\alpha(g^*)}}{1-\sqrt{\tilde\alpha(g^*)}}-\frac{\tilde\alpha(g^*)}{2} 
\left(\frac{2\omega_r}{\Delta}+1\right)
&=\frac{2\omega_r}{\Delta}-1\;,
\ee
where $\tilde\alpha(g)=(2g/\omega_r)^2$.
Expanding $g^*=\omega_r/2-x$ and retaining terms up to second order in $x$, we find 
\bes\label{gstar}
g^*\simeq \frac{\omega_r}{8\omega_r/\Delta+4}\left(\frac{\omega_r}{\Delta}+\frac{3}{2}+\sqrt{\frac{9\omega_r^2}{\Delta^2} - \frac{5\omega_r}{\Delta}-\frac{15}{4}}\right)\;.
\ee
The vertical arrows in Fig.~\ref{fig_JR_vs_g}(a) and the dashed line in Fig.~\ref{fig_JR_vs_g}(b) mark the values of $g^*$ given by this approximate solution. Note that the expression is independent of the temperature, although for high temperatures the TLS truncation of the QRM is inappropriate.\\

\subsection{Current and rectification at finite bias}
\label{sec_IRvseps}

\begin{figure}[ht!]
\begin{center}
\includegraphics[width=0.41\textwidth,angle=0]{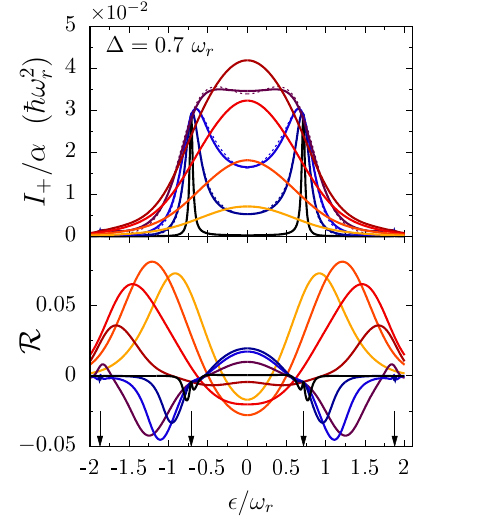}\hspace{-2.2cm}
\includegraphics[width=0.41\textwidth,angle=0]{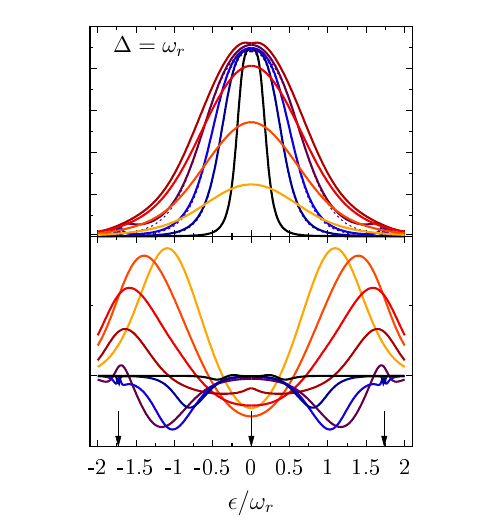}
\hspace{-2.3cm}
\includegraphics[width=0.41\textwidth,angle=0]{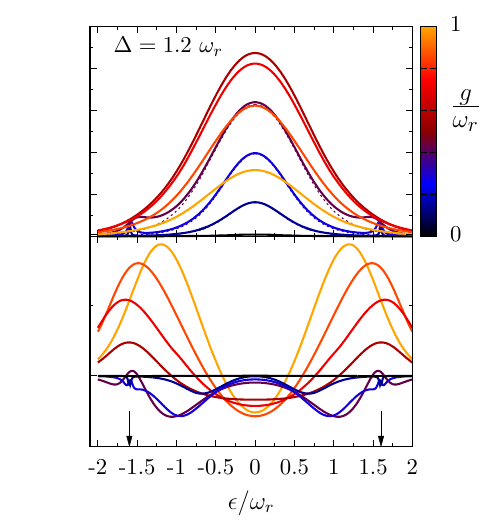}
\caption{\small{Signatures of transition to the USC regime by tuning the bias.
Heat current and rectification \emph{vs.} qubit bias $\epsilon$ for $\Delta/\omega_r=0.7,1$ and, $1.2$ and the same values of the qubit-resonator coupling strength $g$ as in Fig.~\ref{fig_JR_vs_Delta}. Other parameters are as in Fig.~\ref{fig_JR_vs_Delta}.
For $g\ll \omega_r$ the current has peaks at values of bias where the resonance condition $\omega_q=\sqrt{\Delta^2+\epsilon^2}=n\omega_r$ is met. In the USC regime the current displays a single, zero-bias maximum. Upon increasing the coupling strength, the rectification  makes a transition from positive to negative for $\omega_q<\omega_r$, see also Fig.~\ref{fig_JR_vs_g}, and vice-versa for $\omega_q>\omega_r$. The dashed lines  for $g/\omega_r\leq 0.2$  in the upper panels are the analytical results from 2nd-order Van Vleck perturbation theory in $g$, see~\cite{Magazzu2024PRB,Magazzu2025}, and truncation to a four (dressed)-state system. Van Vleck perturbation theory does not capture the change of convexity of the current at zero-bias upon increasing $g$.}}
\label{fig_JR_vs_eps}
\end{center}
\end{figure}

When varying the bias on the qubit, the transition to the USC regime of qubit-resonator coupling has signatures both in the current and the rectification. In Fig.~\ref{fig_JR_vs_eps}, we show the forward current and $\mathcal{R}$ as a function of the bias for the same values of the coupling $g$ considered in Fig.~\ref{fig_JR_vs_Delta}. The  three panels of Fig.~\ref{fig_JR_vs_eps} display the results for  $\Delta/\omega_r<1$, $\Delta/\omega_r=1$, and $\Delta/\omega_r>1$ corresponding to negative detuning, resonance and positive detuning at zero bias, respectively. The resonance condition is attained, for finite bias, when the latter satisfies $\epsilon_{\rm res}=\pm\sqrt{(n\omega_r)^2-\Delta^2}$. In particular, for $\Delta>\omega_r$ only multi-photon resonances ($n>1$) are possible. Consider first the  case where $\Delta<\omega_r$. Similarly to what happens to the thermal conductance, described in~\cite{Magazzu2025}, upon increasing $g$ the current has a transition from a resonant peak regime, where it is suppressed except around resonance, to a zero-bias peak in the USC. The dotted lines show calculations  using second-order Van-Vleck perturbation theory in $g$ for the QRM~\cite{Magazzu2025}: These do not capture this single maximum at $\epsilon=0$ which indicates that perturbation theory in $g$ breaks down. The signature of transition to the USC regime in the rectification is a flip in the sign of $\mathcal{R}$: While at weak to intermediate $g$ the rectification turns from positive to negative as a function of the bias, the opposite occurs in the USC where $\mathcal{R}$ becomes large and positive for large bias.\\
\onecolumngrid

\begin{figure}[ht!]
\begin{center}
\includegraphics[width=0.42\textwidth,angle=0]{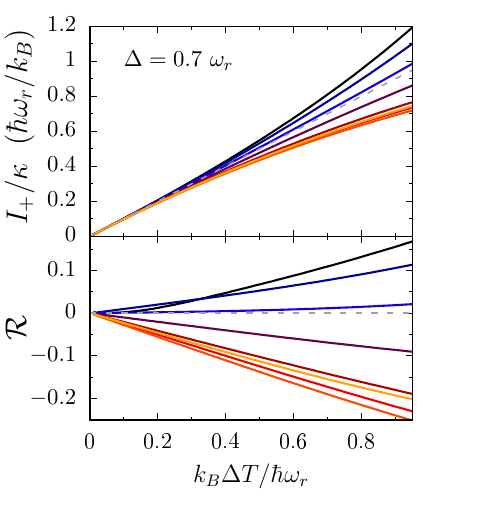}
\hspace{-2.2cm}\includegraphics[width=0.42\textwidth,angle=0]{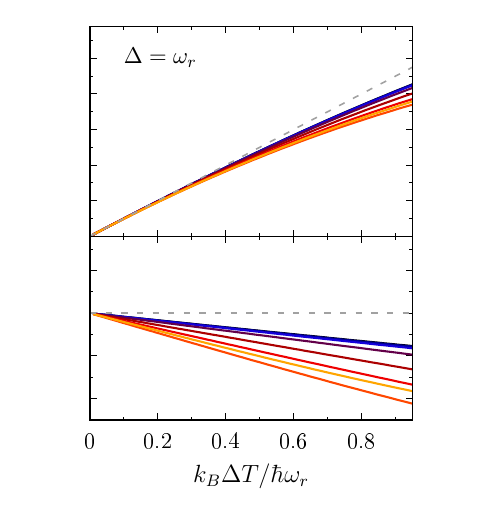}
\hspace{-2.2cm}\includegraphics[width=0.42\textwidth,angle=0]{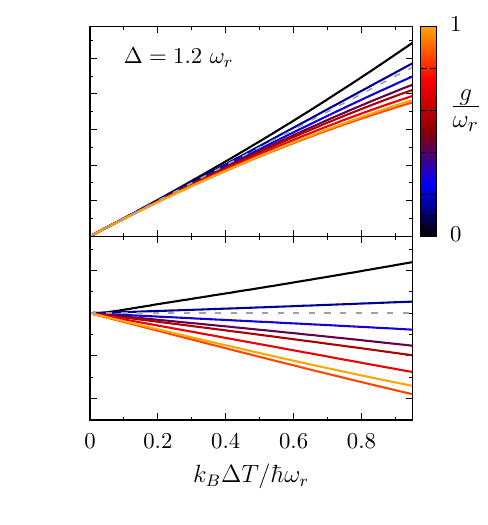}
\caption{\small{Superlinear to sublinear turnover in the current and change of sign of rectification. Current, scaled with the linear conductance $\kappa$, and rectification \emph{vs.} the temperature bias $\Delta T$, for $T=\hbar\omega_r/k_B$ and three values of the detuning at zero bias.
Off-resonance  (left and right panels), the current displays, upon increasing the qubit-resonator coupling $g$, a transition from a superlinear to a sublinear growth with the temperature bias. Correspondingly, the  rectification changes sign. At resonance (center) the scaled current grows always sublinearly and the rectification is negative. 
The  values of the qubit-resonator coupling $g$ are the same as in Figs.~\ref{fig_JR_vs_Delta},
 and ~\ref{fig_JR_vs_eps}. Other parameters are as in Fig.~\ref{fig_JR_vs_Delta}. 
 }}
\label{fig_JR_vs_dT}
\end{center}
\end{figure}
\twocolumngrid
\indent A qualitative account of the results in the weak coupling regime $g/\omega_r\ll 1$ is provided by the RWA, see Appendix~\ref{sec_RWA}, with the further approximation of a two-level system description of the QRM. Equation~\eqref{Q_RWA} gives for the matrix elements of the system coupling operators in the ground and first excited states 
\begin{equation}
\begin{aligned}\label{RWA_u_v}
Q_{R01} &=\;\frac{\Delta}{\omega_q}\frac{\delta - \sqrt{\delta ^2 + 4g_x^2}}{\sqrt{\left(\delta - \sqrt{\delta^2+4g_x^2}\right)^2+4g_x^2}}\;,\\
Q_{L01} &=\;\frac{-2g_x}{\sqrt{\left(\delta - \sqrt{\delta^2+4g_x^2}\right)^2+4g_x^2}}\;,
\end{aligned}
\end{equation}
with $\delta:=\omega_q-\omega_r$ and $ g_x:= g\Delta/\omega_q$.
Far from resonance, $|\delta|\gg g$, we have $Q_{l01}\sim 0$
which suppresses the current, whereas at resonance, $\delta=0$,  we have $Q_{L01}=(\omega_q/\Delta) Q_{R01}=-1/\sqrt{2}$ giving a peak in the current and nonzero (negative) rectification at finite bias because $|Q_{R01}|<|Q_{L01}|$, see Eq.~\eqref{chi}. Note that, within the RWA, this result is independent of $g$, which  accounts for the pinning of the heat current at the resonance condition for different (not too large) values of $g$.
For $\Delta=\omega_r$, the resonance condition is met at $\epsilon=0$ which entails a $g$-independent, zero-bias peak in the current. For $\Delta>\omega_r$, weak qubit oscillator coupling gives small current peaks corresponding to the two-photon resonances and a vanishing current elsewhere. The current becomes substantially different from zero at large coupling and presents a zero-bias peak. The rectification is positive, at small $g$ only for negative detuning $\omega_q<\omega_r$.\\

\subsection{USC effects in transport in the nonlinear transport regime}
\label{sec_IRvsTDT}

Finally, we study the temperature dependence of the transport properties at zero qubit  bias and
for very small $\alpha$ usinng the FSME.
We use the values of qubit frequency splitting $\Delta$ and qubit-resonator coupling $g$ as in Fig.~\ref{fig_JR_vs_eps}. \\
\indent In Fig.~\ref{fig_JR_vs_dT} current and rectification are shown as functions of the temperature bias for $k_B T=\hbar\omega_r$. The current is scaled with the thermal conductance $\kappa$,  defined as $\kappa=dI_+/d\Delta T|_{\Delta T=0}$. In linear regime, namely for small $\Delta T$, the rescaled current $I_+/\kappa$ grows linearly with the temperature bias, according to $I_+\simeq \kappa\Delta T$. Deviations from the linear regime emerge for large enough temperature bias. Interestingly, these deviations depend on the qubit-oscillator coupling. Out of resonance (left and right panel of Fig.~\ref{fig_JR_vs_dT}) the current grows super-linearly at weak coupling and, upon increasing the coupling, turns to a sub-linear dependence on $\Delta T$. Correspondingly the rectification is positive and grows with $\Delta T$ at weak coupling, becomes constant at around $g/\omega_r=0.1$ and grows negatively for stronger couplings. This picture is modified at resonance, $\Delta=\omega_r$. In this case the current scaled with $\kappa$ grows sub-linearly with the temperature bias, independently of the coupling strength, and the rectification is always negative. Beside the sign of $R$, we find in all three panels that $\mathcal{R}$ grows with $\Delta T$ accordingly to what what was found in~\cite{Liu2024} for two coupled two-level systems.

\subsection{The role of steady-state coherences}
\label{reole_ss_coherences}

\begin{figure}[ht!]
\begin{center}
\includegraphics[width=0.45\textwidth,angle=0]{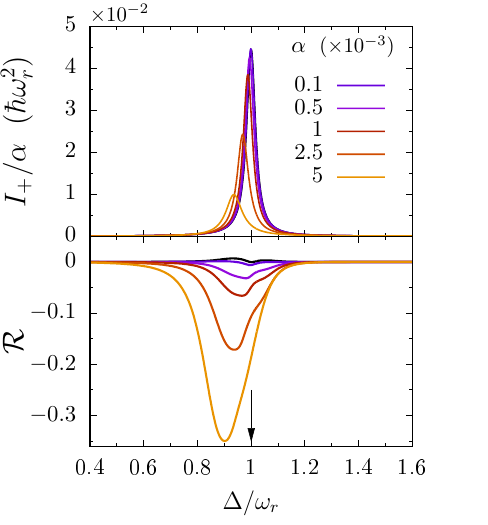}
\caption{\small{Effects of coherence at zero bias.
Heat current and rectification \emph{vs.} qubit splitting $\Delta$, calculated with the PSME, at zero bias for $g/\omega_r=0.01$ and different values of $\alpha$. Other parameters are as in Fig.~\ref{fig_JR_vs_Delta}. The arrow points at the qubit-oscillator resonance condition. The black lines reproduce the results form the FSME: They coincide with the corresponding black lines in Fig.~\ref{fig_JR_vs_Delta}.  Note that the results in the full secular approximation are valid at ultraweak coupling to the baths and are independent of $\alpha$, when the current is scaled with $\alpha$. However when the coupling is such that the rates are comparable with the frequency separation of the quasi-degenerate doublets, $\omega_{21}$ and $\omega_{43}$, see Fig.~\ref{fig_spectrum}, the partial secular approximation is used and the results depend on $\alpha$, which is to be understood as a finite coupling effect.}}
\label{fig_JR_vs_Delta_coh}
\end{center}
\end{figure}

\begin{figure}[ht!]
\begin{center}
\includegraphics[width=0.45\textwidth,angle=0]{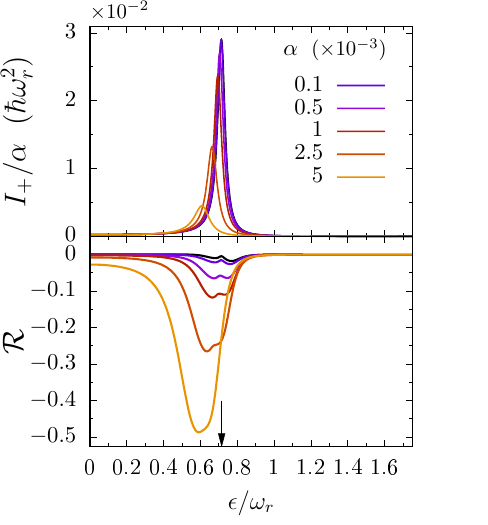}
\caption{\small{Effects of coherence at finite bias.
Current and rectification \emph{vs.} qubit bias $\epsilon$, calculated with the partial secular ME, for $\Delta/\omega_r=0.7$ and $g/\omega_r=0.01$, cf. Fig.~\ref{fig_JR_vs_Delta_coh}.  Other parameters as in Fig.~\ref{fig_JR_vs_eps}. 
The arrow points at the qubit-oscillator resonance condition. The black lines reproduce the results form the FSME and coincide with the corresponding black lines in Fig.~\ref{fig_JR_vs_eps}, left panel (the one in the upper panel overlaps with the PSME result at $\alpha=10^{-4}$). The Rabi model is truncated to the first five levels and the coherences $\rho_{12}$ and $\rho_{34}$, i.e. the ones corresponding to the quasi-degenerate transitions $\omega_{12}$ and $\omega_{34}$, are retained, see Fig.~\ref{fig_spectrum}.}}
\label{fig_JR_vs_eps_coh}
\end{center}
\end{figure}

\indent In Fig.~\ref{fig_JR_vs_Delta_coh} we assess the effects of steady-state coherences on the current and rectification as a function of the qubit splitting at zero bias. We consider a (converged) five-level system truncation of the full quantum Rabi model and  use the PSME~\eqref{BR_partial_secular} to show the results for $g/\omega_r=0.01$. In this coupling regime the spectrum presents, around resonance, two quasi-degenerate doublets, see Fig~\ref{fig_spectrum}(a). Different values of $\alpha$ are considered. For very weak $\alpha$, the result for the current coincides with that of the FSME, Fig.~\ref{fig_JR_vs_Delta}, which is given here for reference by the black curve. However, by increasing $\alpha$, the current scaled with the system-bath coupling, $I_+/\alpha$, which in the PSME is not independent of $\alpha$, changes qualitatively: The current is suppressed by the presence of steady-state coherences and the current peak is shifted from resonance towards $\Delta<\omega_r$. Correspondingly, the rectification is largely enhanced. Suppression of the current by steady-state coherences was found in~\cite{Wang2019} for a junction formed of two coupled qubits and in~\cite{Ivander2022} for a qutrit in the $V$-system configuration.\\  
\indent Similarly to the study as a function of $\Delta$ at zero qubit bias shown in Fig.~\ref{fig_JR_vs_Delta_coh}, we consider in Fig.~\ref{fig_JR_vs_eps_coh} the effect of steady-state coherences on current and rectification as a function of the qubit bias. Also in this case we consider the weak qubit-resonator coupling case $g/\omega_r=0.01$ which displays, around resonance, quasi-degeneracies in the doublets of excited states. The current is calculated within the PSME for negative detuning at zero bias $\Delta/\omega_r=0.7$. The latter reproduces the results of the FSME at very weak $\alpha$ and deviates from it as the system-bath coupling is increased. Also in this case, $I_+/\alpha$ depends, at finite system-bath coupling, on $\alpha$. The current peaks located at resonance  for vanishing $\alpha$, are suppressed and shifted towards $|\epsilon|<|\epsilon_{\rm res}|$ to an extent which depends on $\alpha$.\\
\indent  Note that the present treatment corresponds to the global master equation~\cite{Cattaneo2019} approach and does not neglect the Lamb shift.  Beside describing the effects on transport of finite system-bath coupling, this type of analysis is useful to assess for which values of the system-bath coupling the FSME is appropriate. 
In Appendix~\ref{3LSRWA}, we evaluate the current analytically in the same parameter regimes as in Fig.~\ref{fig_JR_vs_Delta_coh} with a three-level system truncation of the QRM. We find that, while at very small $\alpha$ the results match qualitatively the ones given here at larger values of the system-bath coupling, the contribution of the higher doublet of excited states changes qualitatively the results. 
In particular, while the three-level system truncation accounts for the suppression of the peak current, it does not capture the shift of the peak and fails to reproduce the qualitative features of the rectification (not shown). \\

\section{Experimental implementation}
\begin{figure}[ht!]
\begin{center}
\includegraphics[width=0.45\textwidth,angle=0]{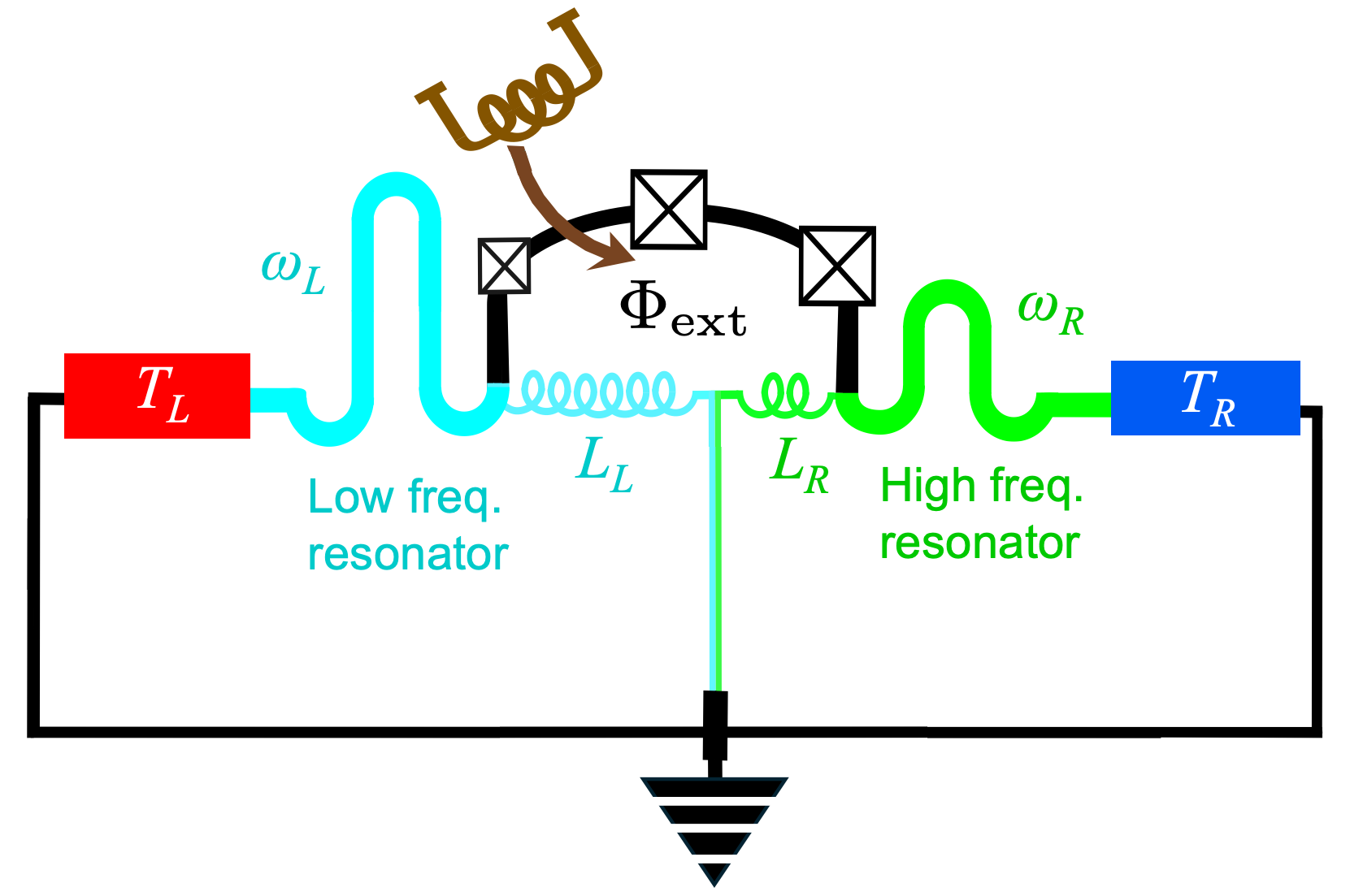}
\caption{\small{Schematic of the experimental implementation of the heat rectification setup based on a superconducting circuit. A flux qubit (central part, with crossed boxes representing the Josephson junctions) is galvanically coupled via two $\lambda/2$ resonators (light blue and green) of different frequencies to metallic heat reservoirs. The qubit is biased via an externally applied magnetic flux $\Phi_{\rm ext}$. The two oscillator frequencies should be very different in order to achieve the QRM setup of Fig.~\ref{fig_setup}. Figure adapted from~\cite{Upadhyay2025}.}}
\label{fig_scheme_exp}
\end{center}
\end{figure}
A candidate device to display the strong qubit-resonator coupling effects on current and rectification shown here is a variation of the one realized in~\cite{Upadhyay2025}.
The setup, schematized in Fig.~\ref{fig_scheme_exp}, consists of a superconducting flux qubit galvanically coupled to two $\lambda/2$ resonators with coupling strengths $g_l$, where $l=L,R$. The latter are proportional to the values of the inductances $L_l$~\cite{Yoshihara2017}. Galvanic coupling has been shown to yield coupling strengths well into the USC regime~\cite{Yoshihara2017,Yoshihara2018}. The resonators are, in turn, connected to metallic heat reservoirs whose temperature is controlled by voltage-biased normal metal–insulator–superconductor (NIS) junctions.  
The flux qubit is biased by an applied magnetic flux  $\Phi_{\rm ext}$ producing the bias $\epsilon$ via the relation $\hbar\epsilon= 2I_p (\Phi_{\rm ext}-\Phi_0/2)$, where $I_p$ is the qubit persistent current and $\Phi_0=h/2e$ is the flux quantum. 
The configuration used in the experiment in~\cite{Upadhyay2025} is symmetric, namely the resonator frequencies $\omega_l$ and their coupling strengths to the qubit are the same in the left and right arm of the device, which does not allow for heat current rectification. Nevertheless, by modifying the frequency $\omega_R$ of the right resonator so that its value is much larger than the other relevant frequency scales, $\omega_R\gg\omega_L,\Delta,g_l$, one can achieve a configuration similar to the one in Fig.~\ref{fig_setup}, with the junction described according to Eq.~\eqref{H_Rabi}. Indeed, in this case, the degrees of freedom of the right resonator are \emph{frozen} and it simply serves as coupler. As a result, one can model the right arm of the setup as a ohmic heat bath directly connected to the qubit with qubit-bath coupling $\alpha_R=\alpha_R(g_R)$ and qubit-resonator coupling $g=g_L$. While $g_R$ should be chosen so as to yield $\alpha_R\ll 1$, the coupling to the left resonator can be designed to give $g_L$ in the range from weak to ultrastrong. Finally, we note that the experimental results in~\cite{Upadhyay2025} for the heat current as a function of the applied bias display a peak at zero applied bias and two side peaks corresponding to the condition $\omega_q=\omega_l$, with $\Delta<\omega_l$ and $g_l\sim0.1~\omega_l$. This is in contrast with our evaluations where, for $\Delta<\omega_r$, the zero-bias peak is absent in similar coupling regimes, see Fig.~\ref{fig_JR_vs_eps}.
However, this discrepancy can be attributed to the symmetric configuration in the experiments. Indeed, using Eq.~\eqref{Jp_TLS} for a qubit interacting with two effective, structured baths with low-frequency ohmic behavior~\cite{Garg1985}, one finds that this symmetry implies that the bias $\epsilon$ appears only in the Bose-Einstein functions $n_l(\omega)$. On the other hand, as shown above, the matrix elements of the coupling operators are crucial to give the resonant peaks and the out-of-resonance suppression of the heat current and they appear explicitly in the asymmetric configuration with a single oscillator.

\section{Conclusions}
\label{conclusions}

We have studied extensively the thermal transport properties of a qubit-resonator system placed between heat baths. The junction is modeled according to the quantum Rabi model and we have considered an implementation with superconducting circuits. These allow for tuning the parameters of the individual components and for reaching the ultrastrong coupling regime. Assuming weak coupling to the heat baths, we found signatures of this nonperturbative regime of qubit-resonator coupling both in the heat current and in the rectification. The latter quantifies the asymmetry in the net current flowing from the hot to the cold bath upon reversing the temperature bias between the two. A setup displaying thermal rectification is on technological interest because it operates as the thermal analog of a diode, i.e. is has a preferential temperature configuration for the heat current and suppresses the current in the opposite configuration. Increasing the qubit-resonator coupling, the rectification changes sign, implying a change in the direction of suppressed current. This occurs generally for negative detuning, namely when the qubit frequency is larger than the one of the resonator. A noticeable suppression of the current, accompanied by a substantial enhancement of rectification, is found in the parameter regimes that allow for nonvanishing steady-state coherences. Finally, we propose a realistic implementation based on an experiment on heat transport in strongly-coupled qubit-resonator systems.\\
\indent Our findings describe the performance and the flexibility of an implementation of the archetypal model of light-matter interaction based on superconducting circuits operating as heat valve and rectifier.

\section{Acknowledgements}

The authors thank Rosario Fazio and Fabio Taddei for discussions and comments. 
LM and MG acknowledge financial support from BMBF (German Ministry for Education and Research), Project No. 13N15208, QuantERA SiUCs, and CRC 1277. The research is part of the Munich Quantum Valley, which is supported by the Bavarian state government with funds from the Hightech Agenda Bavaria. EP acknoweldges financial support from PNRR MUR project
PE0000023-NQSTI, from COST ACTION SUPQERQUMAP, CA21144, and PIACERI, Theoretical Condensed Matter for Quantum Information.

\appendix

\onecolumngrid

\section{Rotating-wave approximation (RWA)}
\label{sec_RWA}

The quantum Rabi Hamiltonian, Eq.~\eqref{H_Rabi}, expressed in the qubit energy basis $\{\ket{g},\ket{e}\}$, assumes the form 
\be{HRabi_energy}
H_{\rm Rabi}=& -\frac{\hbar}{2}\omega_q{\sigma}_z +\hbar \omega_r a^\dagger a + \hbar\left(g_z{\sigma}_z -g_x{\sigma}_x  \right) \left(a^\dagger + a \right)
\ee
with $\omega_q:=\sqrt{\Delta^2+\epsilon^2}$ and 
$g_z= g \epsilon/\omega_q$ and $g_x= g \Delta/\omega_q$.
In this basis, the Pauli matrices read  $\sigma_z =\ket{g}\bra{g}-\ket{e}\bra{e}$ and $\sigma_x =\ket{g}\bra{e}+\ket{e}\bra{g}=\sigma_++\sigma_-$. 
Within the RWA, we retain the excitation-preserving terms in the qubit-oscillator coupling Hamiltonian and neglect the contributions $\propto \sigma_+ a^\dag$ and $ \sigma_- a$ in Eq.~\eqref{H_Rabi}. This amount to approximate the Rabi model with the Jaynes-Cummings Hamiltonian~\cite{Jaynes1963,Shore1993}. The latter admits a simple analytical diagonalization resulting in
\begin{equation}
\begin{aligned}
\label{JCMspectrum}
\omega_0=&-\omega_q/2\;,\\
\omega_{2n-1}=& \left(n-\frac{1}{2} \right)\omega_r - \frac{1}{2}\sqrt{\delta^2 + n 4 g_x^2}\;, \\
\omega_{2n}=& \left(n-\frac{1}{2} \right)\omega_r + \frac{1}{2}\sqrt{\delta^2 + n 4g_x^2}
\end{aligned}
\end{equation}
($n\geq 1$), with the detuning $\delta$ defined as $\delta:=\omega_q-\omega_r$. The corresponding eigenstates, in the energy basis of the uncoupled system, are
\begin{equation}
\begin{aligned}
\label{JCMeigenstates}
|0\rangle &= |{\rm g},0\rangle\;,\\
|2n-1\rangle &= {\rm u}_{n}^-|{\rm e},n-1\rangle + {\rm v}_{n}^-|{\rm g},n\rangle\;,\\
|2n\rangle &= {\rm u}_{n}^+|{\rm e},n-1\rangle + {\rm v}_{n}^+|{\rm g},n\rangle\;,
\end{aligned}    
\end{equation}
with coefficients
\begin{equation}
\begin{aligned}\label{QLR_RWA}
{\rm u}_n^\pm &:=\;\frac{\delta \pm \sqrt{\delta ^2 + n4g_x^2}}{\sqrt{\left(\delta \pm \sqrt{\delta^2+n4g_x^2}\right)^2+n4g_x^2}}\;,\qquad{\rm and}\qquad
{\rm v}_n^\pm :=\;\frac{-\sqrt{n}2g_x}{\sqrt{\left(\delta \pm \sqrt{\delta^2+n4g_x^2}\right)^2+n4g_x^2}}\,.
\end{aligned}
\end{equation}

Within the RWA, the relevant matrix elements of the system's coupling operators ($\hat{Q}_L=\hat{a}+\hat{a}^\dag$ and $\hat{Q}_R=\sigma_z\epsilon/\omega_q-\sigma_x \Delta/\omega_q$, in the qubit energy basis) read, in the subspace $n\leq 2$,
\be{Q_RWA}
Q_{L01}&={\rm v}_1^-\;,\qquad Q_{R01}={\rm u}_1^-\frac{\Delta}{\omega_q}\\
Q_{L02}&={\rm v}_1^+\;,\qquad Q_{R02}={\rm u}_1^+\frac{\Delta}{\omega_q}\\
Q_{L12}&=0\;,\qquad Q_{R,12}=({\rm v}_1^-{\rm v}_1^+-{\rm u}_1^-{\rm u}_1^+)\frac{\epsilon}{\omega_q}\\
Q_{L00}&=0\;,\qquad Q_{R00}=-\frac{\epsilon}{\omega_q}\\
Q_{L11}&=0\;,\qquad Q_{R11}=[({\rm v}_1^-)^2-({\rm u}_1^-)^2]\frac{\epsilon}{\omega_q}\\
Q_{L22}&=0\;,\qquad Q_{R22}=[({\rm v}_1^+)^2-({\rm u}_1^+)^2]\frac{\epsilon}{\omega_q}\;.
\ee

\section{Caldeira-Leggett model}
\label{appendix_CLM} 
The Caldeira-Leggett model~\cite{Caldeira1981,Caldeira1983} describes an open system bilinearly interacting via the operator $\hat X$ with a heat bath of harmonic oscillators
\be{H_Caldeira_Leggett}
\hat H=& \hat H_{\rm S}+
\hat{H}_{B}+\hat{H}_{SB}\\
=&\hat H_{\rm S}+\frac{1}{2}\sum_{j=1}^{N} \left[
\frac{\hat{p}^{2}_{j}}{m_{j}}+m_{j}\omega^{2}_{j}\left ( \hat{x}_{j}
-\frac{c_{j}} {m_{j}\omega^{2}_{j} } \hat{X} \right )^{2}\right].
\ee
The bath and its interaction with the system are characterized by the spectral density function $J(\omega)$ defined as
\begin{equation}\label{J}
J(\omega)=\frac{\pi}{2}\sum_{j=1}^{N}\frac{c_{j}^{2}}{m_{j}\omega_{j}}\delta(\omega-\omega_{j})\;.
\end{equation}
Let us introduce the coupling constants with dimension of an angular frequency 
\[
\lambda_j=\frac{x_0}{\sqrt{2\hbar m_j \omega_j}}c_j\;.
\]
In the contest of the spin-boson model, it is customary to define the modified spectral density function~\cite{Weiss2012} 
\be{G}
G(\omega):=\sum_{j=1}^{N}\lambda_j^2\delta(\omega-\omega_{j})=\frac{x_0^2}{\pi\hbar}J(\omega)\;.
\ee
For a harmonic oscillator of frequency $\omega_r$ coupled to a strictly ohmic bath (quantum Brownian motion), $J(\omega)=M\gamma\omega$, where $\gamma$ is the memoryless friction kernel and $x_0=\sqrt{\hbar/2M\omega_r}$.

Since a bosonic bath $\hat{x}_j=\sqrt{\hbar/2m_j\omega_j}(b_j+b_j^\dag)$, defining the \emph{dimensionless} system position operator $\hat{Q}$ via $\hat{X}=x_0\hat{Q}$, we can write
\be{HCL}
\hat H=& \hat H_{\rm S}+\sum_j\hbar\omega_j b_j^\dag b_j -\hat{Q}\sum_j x_0c_j\sqrt{\frac{\hbar}{2m_j\omega_j}}(b_j+b_j^\dag)+\hat A^2\\
=&\tilde{H}_{\rm S}+\sum_j\hbar\omega_j b_j^\dag b_j -\hat{Q}\hat B\;,
\ee
where 
\be{Hterms}
\tilde{H}_{\rm S}:=&\;\hat H_{\rm S}+\hat A^2\qquad{\rm and}\qquad  \hat{B}:=\sum_j \hbar\lambda_j(b_j+b_j^\dag)\;.
\ee
Here, $\tilde{H}_{\rm S}$ is renormalized by absorbing the term $\hat A^2$, with the dimension of an energy, which is proportional to the square of system operator coupled to the bath and is given by, see Eqs.~\eqref{H_Caldeira_Leggett}-\eqref{G},
\begin{equation}\label{A2}
\hat A^2=\sum_{j=1}^{N}\frac{c_{j}^{2}}{2 m_{j}\omega_{j}^2}\hat{X}^2=\frac{1}{\pi}\int_{0}^{\infty}d\omega\frac{J(\omega)}{\omega}\hat{Q}^2x_0^2 = \hbar\int_{0}^{\infty}d\omega\frac{G(\omega)}{\omega}\hat{Q}^2\equiv\mu \hat Q^2 \;.
\end{equation}
For a ohmic-Drude spectral density function $G(\omega)=\alpha\omega/[1+(\omega/\omega_c)^2]$, we have $\mu=(\pi/2)\alpha\hbar\omega_c $.  
Note that if the system operator coupling to the bath is the qubit's Pauli operator, $\hat{Q}=\sigma_z/2$, then $\hat{Q}^2$ is proportional to the identity and the term $\hat A^2$ merely constitutes a constant shift in the Hamiltonian.\\

\section{Bath correlation function and power spectral density}
\label{appendix_BCF}

The correlation function of the baths operators $\hat{B}_l:=\sum_{ j}\hbar\lambda_{l j}(b_{l j}+ b_{l j}^\dag)$, where the time evolution is induced by the free bath Hamiltonian, is given by
$$\langle \hat{B}_l(t)\hat{B}_l(0)\rangle=\hbar^2\int_0^\infty d\omega G_l(\omega)\left[\coth\left(\frac{\beta_l\hbar\omega}{2}\right)\cos(\omega t)-{\rm i}\sin(\omega t)\right]\;.$$
The functions $W_{lnm}$ are the one-sided Fourier transform of the above quantity, namely
\be{}
W_{l nm}=&\int_{0}^{\infty}dt\langle \hat{B}_l(t)\hat{B}_l(0)\rangle 
e^{-\ii\omega_{nm} t}
\;.
\ee
Since real and imaginary parts of the bath correlation function are even and odd function of $t$, respectively, the power spectral density is the real function $S_{l}(\omega_{nm})=\int_{-\infty}^{\infty}dt\langle \hat{B}_l(t)\hat{B}_l(0)\rangle 
e^{\ii\omega_{nm} t}=2{\rm Re}W_{lmn}$.\\
\indent For the sake of readability, we give the explicit result for $W_{lnm}$ skipping the bath index $l$. It is understood that temperature, coupling, and spectral density function depend on the bath considered. For a ohmic-Drude bath correlation function with cutoff frequency $\omega_c$, namely  $G(\omega)=\alpha\omega/[1+(\omega/\omega_c)^2]$, the functions $W_{nm}$ can be explicitly evaluated and read~\cite{Magazzu2024PRB}
\be{Wnm_explicit}
\frac{W_{nm}}{\hbar^2} 
&=\pi G(\omega_{nm})n(\omega_{nm})
-\ii\frac{\pi}{2}G(\omega_{nm})\left[\cot\left(\frac{\beta\hbar\omega_c}{2}\right)+\frac{\omega_c}{\omega_{nm}}\right]+\ii\frac{2\pi\alpha\omega_c^2}{\hbar\beta}\sum_{k=1}^\infty
\frac{\nu_k \omega_{nm}}{(\omega_c^2-\nu_k^2)(\omega_{nm}^2+\nu_k^2)}\;,
\ee
where we used $G(-\omega)=-G(\omega)$ for the real part $W'_{nm}$ and where $\nu_n:=2\pi n/\beta\hbar$ are the Matsubara frequencies. Since the spectral density function is an odd function of the Bohr frequency
\be{}
W'_{nm}=\hbar\pi \alpha\frac{|\omega_{nm}|}{1+(\omega_{nm}/\omega_c)^2} \begin{cases}
n(|\omega_{nm}|)    \qquad\qquad n>m\\
n(|\omega_{nm}|)+1 \qquad n<m
\end{cases}\;.
\ee 
At small frequencies $G(\omega)\sim \alpha\omega$, thus we have
\be{}
W_{nn}=\lim_{\omega_{nm}\to 0}\frac{W_{nm}}{\hbar^2}=\frac{\pi\alpha}{\hbar\beta}-\ii\frac{\pi}{2}\alpha\omega_c\;.
\ee
Taking $G(\omega)\simeq \alpha\omega$, the series expansion of $\cot(x)$ yields for the imaginary part
\be{}
\frac{W''_{nm}}{\hbar^2}\simeq\frac{2\pi\alpha\omega_c^2}{\hbar\beta}\sum_{k=1}^\infty
\frac{\nu_k \omega_{nm}}{(\omega_c^2-\nu_k^2)(\omega_{nm}^2+\nu_k^2)}
-\frac{2\pi\alpha\omega_c}{\hbar\beta}\sum_{k=1}^\infty
\frac{\omega_{nm}}{(\omega_c^2-\nu_k^2)}-\frac{\pi\alpha\omega_{nm}}{\hbar\beta}-\frac{\pi\alpha\omega_c}{2}\;.
\ee
In the large temperature limit ($\nu_1>\omega_{nm}$)
\be{ImWapprox}
\frac{W''_{nm}}{\hbar^2}\simeq\frac{2\pi\alpha\omega_c}{\hbar\beta}\sum_{k=1}^\infty
\frac{\omega_{nm}}{(\omega_c+\nu_k)\nu_k}
-\frac{\pi\alpha\omega_{nm}}{\hbar\beta}-\frac{\pi\alpha\omega_c}{2}\simeq \alpha\omega_{nm} -\frac{\pi\alpha\omega_{nm}}{\hbar\beta}-\frac{\pi\alpha\omega_c}{2}\;.
\ee
We use this approximated form for the imaginary part of $W_{nm}$ for the analytical evaluation of the current shown in Appendix~\ref{3LSRWA}.

\section{Three-level system truncation: Analytical results for weak qubit-oscillator coupling}
\label{3LSRWA}

In this section we provide analytical results for a three-level system (3LS) truncation of the full Rabi model at zero qubit bias, $\epsilon=0$. 
This is a minimal model where steady-state coherences contribute to the current and rectification. Specifically we consider the coupling regime $g\ll \omega_r,\Delta$ 
where the spectrum presents quasi-degenerate excited levels at resonance, $\Delta=\omega_r$, a situation where the PSME is appropriate. In this setting, the quasi degenerate subspace corresponds to the Bohr frequency $\omega_{21}$, see Fig.~\ref{fig_spectrum}, namely the only nonvanishing steady-state coherence is $\rho_{12}$. 
We can use the RWA to describe the qubit-oscillator system. Within this approximation, the matrix elements of $\hat Q_l$ are given  by Eq.~\eqref{Q_RWA} and we obtain for the relevant elements of the Redfield tensor  
\be{}
&\mathcal{K}_{1212}^{(2)}=-\frac{1}{\hbar^2}\sum_{l }\Big\{|Q_{l 01}|^2 W_{l 0 2}+
|Q_{l 02}|^2 W^*_{l 0 1}\Big\}\;,\qquad \mathcal{K}_{1221}^{(2)}=0\;,\\
&\mathcal{K}_{1200}^{(2)}=\frac{1}{\hbar^2}\sum_{l }Q_{l 02}^*Q_{l 10}\left[ W_{l 1 0} + W^*_{l 20}\right]\;,\qquad\mathcal{K}_{0012}^{(2)}=\frac{1}{\hbar^2}\sum_{l }Q_{l 20}^*Q_{l 01}\left[ W_{l 02} + W^*_{l 01}\right]\;,\\
&\mathcal{K}_{1211}^{(2)}=-\frac{1}{\hbar^2}\sum_{l }Q_{l 02}Q^*_{l 01}W^*_{l 01}\;,\qquad\mathcal{K}_{1112}^{(2)}=-\frac{1}{\hbar^2}\sum_{l }Q_{l 01}Q^*_{l 02}W^*_{l 01}\;,\\
&\mathcal{K}_{1222}^{(2)}=-\frac{1}{\hbar^2}\sum_{l }Q^*_{l 01}Q_{l 02}W_{l 02}\;,\qquad\mathcal{K}_{2212}^{(2)}=-\frac{1}{\hbar^2}\sum_{l }Q^*_{l 02}Q_{l 01}W_{l 02}\;,\\
&\Gamma_{nm}:=\mathcal{K}_{nnmm}^{(2)}=\frac{1}{\hbar^2}\sum_{l }|Q_{l nm}|^22{\rm Re}W_{l nm}\;,\\
&\Gamma_{12}=0\;,\qquad\qquad \Gamma_{21}=0\;,\\
&\Gamma_{10}=2\pi\sum_{l }|Q_{l 01}|^2 G_l(\omega_{10})n_l(\omega_{10})\qquad\qquad \Gamma_{01}=2\pi\sum_{l }|Q_{l 01}|^2 G_l(\omega_{10})[n_l(\omega_{10})+1]\;,\\
&\Gamma_{20}=2\pi\sum_{l }|Q_{l 02}|^2 
G_l(\omega_{20})n_l(\omega_{20})\qquad\qquad \Gamma_{02}=2\pi\sum_{l }|Q_{l 02}|^2 G_l(\omega_{20})[n_l(\omega_{20})+1]\;.
\ee
This shows that, even in the limit of vanishing $\omega_{12}$, the corresponding element of the kernel stays finite. 
The solution of Eq.~\eqref{RME} with $\rho_{12}\neq 0$ and setting to zero the other coherences is then 
\be{rho12}
\rho_{12}'&=a_0\rho_{00}+a_1\rho_{11}+a_2\rho_{22}\qquad a_i:=\frac{\mathcal{K}^{(2)''}_{12ii}- \mathcal{K}^{(2)'}_{1212} b_i}{\omega_{12}-\mathcal{K}^{(2)''}_{1212}}\\
\rho_{12}''&=-b_0\rho_{00}-b_1\rho_{11}-b_2\rho_{22}\qquad b_i:=\frac{[\omega_{12}-\mathcal{K}^{(2)''}_{1212}]\mathcal{K}^{(2)'}_{12ii}+\mathcal{K}^{(2)'}_{1212}\mathcal{K}^{(2)''}_{12ii}}{[\omega_{12}-\mathcal{K}^{(2)''}_{1212}]^2+[\mathcal{K}^{(2)'}_{1212}]^2}\;,
\ee
where $\omega:=\omega_{12}-\mathcal{K}^{(2)''}_{1212}$ and $\Omega:=\mathcal{K}^{(2)'}_{1212}
$.

\begin{figure}[ht!]
\begin{center}
\includegraphics[width=0.55\textwidth,angle=0]{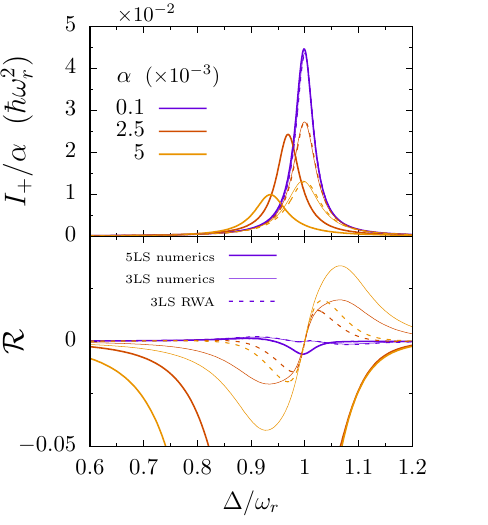}
\caption{\small{Forward current and rectification vs. $\Delta$ at zero qubit bias. Dashed lines: Three-level-system (3LS) truncation with analytical evaluation in the RWA, with $W''_{lnm}$ given by Eq.~\eqref{ImWapprox}. Solid thin lines: Exact treatment (including the bath-renormalized system Hamiltonian $\tilde{H}_{\rm S}$) within the 3LS truncation. Solid thick lines:  exact treatment  within a 5LS truncation, same curves as in Fig.~\ref{fig_JR_vs_Delta_coh}. The three results for the current agree qualitatively for small $\alpha$. For finite $\alpha$ higher levels contribute and the renormalization term $\mu_L \hat{Q}_L^2$ becomes noticeable within the 3LS truncation schemes. The rectification in the 3LS truncation schemes differ qualitatively form the one calculated with the 5LS truncation even at small $\alpha$.}}
\label{fig_3LS}
\end{center}
\end{figure}

The populations are the solution of 
\be{Pop}
0&=\tilde\Gamma_{10} + (\tilde\Gamma_{11}-\tilde\Gamma_{10})\rho_{11}+ (\tilde\Gamma_{12}-\tilde\Gamma_{10})\rho_{22}\\
0&=\tilde\Gamma_{20} + (\tilde\Gamma_{21}-\tilde\Gamma_{20})\rho_{11}+ (\tilde\Gamma_{22}-\tilde\Gamma_{20})\rho_{22}\\
\rho_{00}&=1-\rho_{11}-\rho_{22}\;,
\ee
with $\tilde\Gamma_{ni}:=\Gamma_{ni}+2[
\mathcal{K}^{(2)'}_{nn12}a_i+
\mathcal{K}^{(2)''}_{nn12}b_i]$.
Explicitly
\be{pop_solutions}
\rho_{22}=\frac{A_2+B_2A_1}{1-B_2B_1}\;.\qquad \rho_{11}=A_1+B_1\rho_{22}\;,\qquad \rho_{00}=1-\rho_{11}-\rho_{22}\;,
\ee
where
\be{}
A_1=\frac{\tilde\Gamma_{10}}{\tilde\Gamma_{10}-\tilde\Gamma_{11}}\;,\qquad B_1=\frac{\tilde\Gamma_{12}-\tilde\Gamma_{10}}{\tilde\Gamma_{10}-\tilde\Gamma_{11}}\;,\qquad
A_2=\frac{\tilde\Gamma_{20}}{\tilde\Gamma_{20}-\tilde\Gamma_{22}}\;,\qquad B_1=\frac{\tilde\Gamma_{21}-\tilde\Gamma_{20}}{\tilde\Gamma_{20}-\tilde\Gamma_{22}}\;.
\ee
Finally, observing that $Q_{lnm}\in\mathbb{R}$, see Eq.~\eqref{QLR_RWA}, the current to bath $r$ reads, according to Eq.~\eqref{I2},
\be{current3LS}
I_r=&-\frac{2}{\hbar^2}\Big\{\left[Q_{r01}^2\bar{W}'_{r10}+Q_{r02}^2\bar{W}'_{r20}\right]\rho_{00}+
Q_{r01}^2\bar{W}'_{r01}\rho_{11}
+
Q_{r02}^2\bar{W}'_{r02}\rho_{22}
\\
&\qquad+
Q_{r02}Q_{r01}\left[\left(\bar{W}'_{r02}+\bar{W}'_{r01}\right)\rho'_{12}-
\left(\bar{W}''_{r02}-\bar{W}''_{r01}\right)\rho''_{12}
\right]\Big\}\\
&=2\pi\alpha\Big\{Q_{r01}^2\omega_{10}^2\theta(\omega_{10})\left[\rho_{11}-n_r(\omega_{10})(\rho_{00}-\rho_{11})\right]\rho_{00}
\\&\qquad+
Q_{r02}^2\omega_{20}^2\theta(\omega_{20})\left[\rho_{22}-n_r(\omega_{20})(\rho_{00}-\rho_{22})\right]\rho_{00}
\\
&\qquad+
Q_{r02}Q_{r01}\left[
\omega_{20}^2\theta(\omega_{20})
\left[n_r(\omega_{20})+1\right]
+
\omega_{10}^2\theta(\omega_{10})
\left[n_r(\omega_{10})+1\right]\right]\rho'_{12}\\
&\qquad+
Q_{r02}Q_{r01}\left[
(\omega_{20}^2-\omega_{10}^2)
\left(\frac{1}{\pi}-\frac{1}{\hbar\beta_r}\right)
+(\omega_{20}-\omega_{10})\frac{1}{2}\omega_c\
\right]\rho''_{12}\Big\}
\;,
\ee
where $\theta(\omega)=[1+(\omega/\omega_c)^2]^{-1}$ is the Drude-type cutoff function. To see how the steady-state coherence suppresses the current, one can specialize Eq.~\eqref{current3LS} to the resonant case, $\Delta=\omega_r$,  where $\omega_{10}\simeq\omega_{20}\gg \omega_{21}$. The RWA yields $Q_{R02}=-Q_{R01}$, and we have for the current to bath $R$ the approximate expression
\be{}
I_R\simeq 4\pi\alpha Q_{R01}^2\omega_{10}^2\theta(\omega_{10})\Big\{\left[\rho_{11}-n_R(\omega_{10})(\rho_{00}-\rho_{11})\right]\rho_{00}-
\left[n_R(\omega_{10})+1\right]\rho'_{12}\Big\}\;.
\ee
Since $\rho'_{12}\geq 0$, the second term in parenthesis is subtracted from the first and gives the suppression of the current. Note that in the equivalent situation where the temperature bias is reversed and the current to bath $L$ is calculated, the result is the same because $Q_{L02}=Q_{L01}$ and $\rho'_{12}\leq 0$. Moreover, for a thermal state of the reduced density matrix, the above expression returns vanishing heat current.\\
\indent Figure~\ref{fig_3LS} shows, with dashed lines, the results of the analytical evaluations of the current within the 3LS truncation and RWA, Eqs.~\eqref{rho12} and~\eqref{pop_solutions}-\eqref{current3LS}. For the imaginary part of the functions $W_{lnm}$ we use the approximate evaluation in Eq.~\eqref{ImWapprox}. Also, the bath-induced renormalization of the system Hamiltonian $\mu_L\hat{Q}^2_L$ (since $\hat{Q}^2_R=\mathrm{1}$, the renormalization from the bath on the qubit is a constant shift, see Eq.~\eqref{H_Caldeira_Leggett2baths}) is neglected.
The resulting curves, for different values of system bath coupling $\alpha$, are compared with an exact treatment within a 3LS truncation of the Rabi Hamiltonian and with a converged five-level system (5LS) truncation, as in the main text. For very small $\alpha$, the current curves  given by the three approaches are qualitatively similar, as the renormalization is small and levels higher than the third do not contribute significantly. For finite $\alpha$, however, this is no more the case. While the level renormalization entails a small shift within the 3LS treatments, the contribution from higher levels manifests itself in a qualitative difference in the position and height of the current peaks. Specifically, the 3LS model underestimates the current suppression from steady-state coherences and does not reproduce the large shift in the position of the current peak towards lower frequencies (negative detuning). The rectification predicted by the 3LS truncation schemes differs qualitatively for every $\alpha$ from the 5LS truncation scheme. The latter results coincide with those shown in Fig.~\ref{fig_JR_vs_Delta_coh}. Finally we observe that the RWA is a good approximation for all values of $\alpha$ considered here, indeed when applied to the 5LS truncation, it reproduces the results of the exact renormalization for both the current and the rectification (not shown). As a result we can attribute the qualitative differences in the curves of Fig.~\ref{fig_3LS} to the choice of truncation scheme.

\newpage

\twocolumngrid

%

\end{document}